# A magneto-mechanical gyroscope with spintronic readout

Andrea Meo[1,*], Francesca Garescì[2,*], Pedro Bossi Nuñez[1], Maddalena Fiorentino[1], Teresa Natale[1], Ludovico Dindelli[1], Victor Lopez-Dominguez[3], Banibrato Sinha[4], Riccardo Tomasello[1], Francesco Dell'Olio[1], Pedram Khalili Amiri[4], Mario Carpentieri[1],Giovanni Finocchio[5,*]

[1] *Department of Electrical and Information Engineering, Politecnico of Bari, 70125 Bari, Italy*

[2] *Department of Engineering, University of Messina, I-98166, Messina, Italy*

[3] *Institute of Advanced Materials (INAM), Universitat Jaume I, Castellon, 12006, Spain*

[4] *Department of Electrical and Computer Engineering, Northwestern University, 2145 Sheridan Road, Evanston, 60208, Illinois, USA*

[5] *Department of Mathematical and Computer Sciences, Physical Sciences and Earth Sciences, University of Messina, I-98166, Messina, Italy*

[*]Corresponding authors: andrea.meo@poliba.it, francesca.garesci@unime.it, gfinocchio@unime.it

Gyroscopes are essential elements in navigation, consumer electronics, robotics, and aerospace applications. Most micro electro-mechanical systems (MEMS) implementations rely on capacitive sensing mechanisms, which limit the dimensional scaling to the micrometer scale. In this work, we introduce a MEMS-like two-degree-of-freedom (2-DOF) gyroscope that exploits the rectification functionality of magnetic tunnel junctions (MTJs) as its readout mechanism and as the transducer of the mechanical dynamics. Experimentally characterized MTJs have been used to calibrate and perform an experiment-informed design of the magneto-mechanical model combining micromagnetic theory with 2-DOF mechanical equations. We demonstrated that the output is linear with angular rate, and that the proposed device is able to extract the angular rate in dynamic cases exploiting a homodyne demodulation approach. The results open a path towards a compact, complementary metal-oxide semiconductor (CMOS)-compatible readout pathway that relaxes reliance on tight capacitive gaps and motivates multi-physics designs of the device.

## I. INTRODUCTION

Gyroscopes are core components of inertial measurement units used in navigation, consumer electronics, automotive safety, robotics, and aerospace[1–5]. The dominant commercial solution is the capacitive micro-electro-mechanical (MEMS) vibratory gyroscope[2,5,6], where the Coriolis-induced displacement, typically on the nanometre-scale, of a proof mass along a sense axis is transduced into a measurable variation of capacitance between electrodes. The readout relies on capacitive bridges with synchronous demodulation and, in most high-performance designs, closed loops to restore the driving force. While this architecture is mature, further dimensional scaling is constrained[4] by the requirement for well-controlled capacitive gaps, large proof mass area[7], and low-parasitic interconnects, and performance can degrade in harsh electromagnetic or radiation environments[5].

Spintronics is now a very active branch of electronics that exploits the spin of the electron in addition to its charge[8]. Spintronic devices based on magnetic tunnel junctions (MTJs) [8–10] combine scalability to nanometric dimensions, intrinsic nonvolatility, radiation hardness, and compatibility with standard Complementary Metal-Oxide-Semiconductor (CMOS) processes. MTJs have demonstrated wide applicability[8,9] as memory elements, magnetic sensors, and radiofrequency to direct current (RF-to-DC) rectifiers through the spin-torque diode (STD) effect[11,12]. Mechanical systems integrating spintronic technology offer an alternative path to inertial sensing [13–18]. Initial proposals of spintronic-based accelerometers and gyroscopes employed MTJs as passive elements, as tunnel magnetoresistance (TMR) sensors grown on MEMS substrates to improve the readout signal[19]. There have also been proposals for spintronic accelerometers and gyroscopes based on MTJs [14,15,20,21], which would require custom growth and fabrication processes, resulting in high production costs. More recently, there have been proposals based on coupled MTJs and the STD effect [16–18,22]. These would allow a direct and robust electrical readout of the proof mass motion without capacitance-to-voltage conversion circuits or, alternatively, could be exploited to improve their reading performance with the information provided by the STD output.

Here, we propose a sub-micrometer magneto-mechanical gyroscope where the displacement of a magnetic proof mass induced by the Coriolis response upon rotation modulates the rectified voltage of the MTJs working as STDs placed along the sense axis. This preserves the well-established vibratory gyroscope mechanics, while replacing capacitive plates with a compact spintronic transduction that supports sub-micrometre drive and sense amplitudes, is intrinsically rad-hard, scalable and low-power. Because the STD readout does not rely on sub-100-nm capacitive gaps, the architecture can be scaled down to smaller footprints and to higher sense-mode frequencies without increasing parasitic capacitances and without requiring charge-amplifier front ends. In the mode-matched regime of a vibratory gyroscope, the Brownian-limited noise-equivalent rotation-rate spectral density[23,24] scales as $S_{\Omega}^{1/2} \approx 1/X\sqrt{k_B T/m\omega_s Q_s}$, where $m$ is the effective mass of the proof mass, $\omega_s$ and $Q_s$ are its resonance angular frequency and quality factor, $T$ is temperature, and $X$ is the drive amplitude (applied along the drive axis). Increasing $\omega_s$ and $Q_s$ while maintaining a stable drive amplitude therefore lowers the thermomechanical limit, motivating operation in the high-frequency regime enabled by such a nanoscale spintronic readout.

We first introduce the working principle of the device and discuss the model employed in this study. Then, we present the characterization of the magnetic system. Afterwards, we prove the operation of the proposed devices in the case of a steady-state and a realistic scenario time-dependent angular rate via our modelling approach. Finally, perspectives and conclusions are discussed.

## II. WORKING PRINCIPLE

The mechanical system employs the established vibratory gyroscope paradigm[6,25], such as tuning-fork and dual-mass resonators, with drive ($\omega_d$) and sense ($\omega_s$) modes. Figure 1 summarizes the proposed sensor architecture. A MEMS-like resonator (anchors + springs) provides a driven oscillation along the drive axis (x-axis in a Cartesian reference system) at angular frequency $\omega_d$ with amplitude $X$. Under an applied angular rate $\Omega_z$ about the out-of-plane axis (z-axis), the Coriolis force excites the sense mode, producing a displacement Y along the sense-axis (y-axis). A magnetic proof mass attached to the resonator generates a magnetic stray field (red lines in Fig. 1a) which interacts with the two MTJs placed symmetrically ($d_1 = d_2$ in Fig. 1a) about the proof mass along the sense-axis. Each MTJ works as a STD, at a specific frequency driven by a microwave current. The sense displacement ($d_1 \neq d_2$) modulates the local stray field (red lines) at the two MTJ locations, resulting in a variation of the rectified voltage $V_{\mathrm{dc}}$ that depends on the local field changes. Taking a differential output from the two STDs ($\Delta V_{\mathrm{dc}} = V_{\mathrm{dc},1} - V_{\mathrm{dc},2}$) suppresses common-mode drift and converts the sense displacement into a voltage proportional to $\Omega_z$.

A key point of this proposal, borrowed from typical commercial MEMS-based devices, is the decoupling of the different phenomena involved, which relies on their different characteristic frequencies. In fact, a typical range of the ferromagnetic (FM) resonance frequency is between 100s of MHz and 10s of GHz[26–28], that is orders of magnitude larger than characteristic frequencies of mechanical systems (10s kHz to MHz), such as MEMS membranes, which in turn are larger than those of the physical phenomena to be measured (Hz to kHz). Moreover, the proposed device works in matched-mode operation, i.e. $\omega_d \approx \omega_s$, to maximize the sense response, and we rely on closed-loop force-rebalance[23] to maintain it.

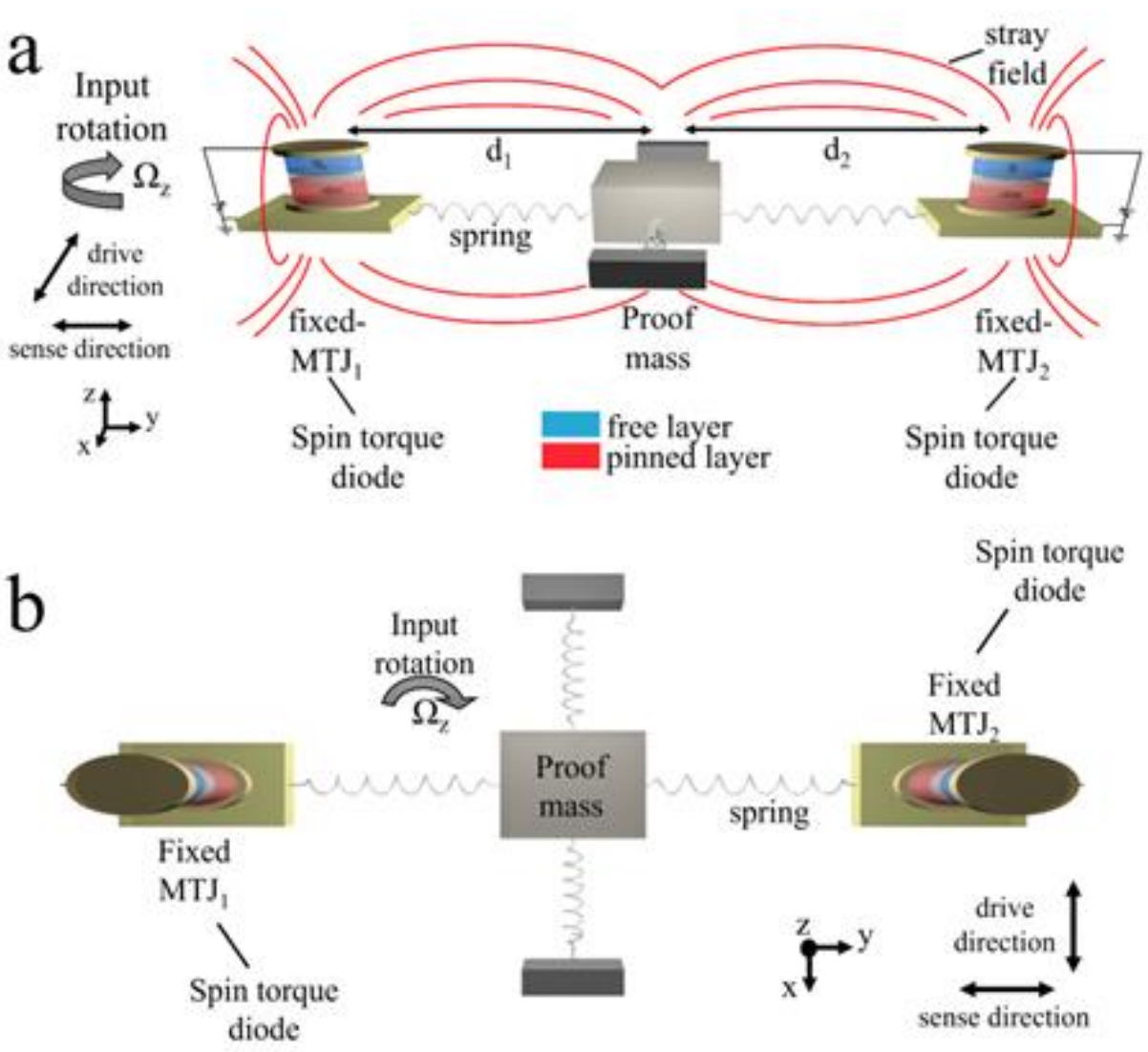

Fig. 1 Sketch of the proposed spintronic gyroscope. (a) Perspective view and (b) top view. The stray field lines have been omitted in the top view for the sake of clarity.

### A. Readout scheme

The readout scheme allows us to map the displacement of the magnetic proof mass linearly onto the output voltage generated by the STDs. This, in turn, is translated into $\Omega_z$ via voltage-displacement calibration. The STD response, which plays a key role in the readout scheme, is based on the rectification response of STDs, where the injection of a small radiofrequency alternating current (ac) into an MTJ can generate a rectified voltage ($V_{\mathrm{dc}}$) across the device when the frequency is at the ferromagnetic resonance [11,12]. Since this response can be tuned by an external magnetic field[29], such as the stray field generated by the magnetic proof mass, the STD effect provides a direct link between displacement of the proof mass induced by the external rotation and $V_{\mathrm{dc}}$.

When we bias the STD response with an external magnetic field $H$, we can define the STD sensitivity ($S$) as the ratio between the voltage variation ($\partial V_{\mathrm{dc}}$) and the change in the field ($\partial H$): $S \equiv \partial V_{\mathrm{dc}}/\partial H$. In our device, $H$ is the stray field generated by the magnetic proof mass, and the strength of this field acting on the MTJ varies as the magnetic proof mass displacement changes due to the driving motion and the Coriolis induced response. Therefore, if we define the field gradient at the MTJ position as $G \equiv \partial H/\partial y$, we can express the differential readout voltage ($\Delta V_{\mathrm{dc}} = V_{\mathrm{dc},1} - V_{\mathrm{dc},2}$), generated by the two STDs as:

$$\Delta V_{\mathrm{dc}} \approx 2SGY. \qquad (1)$$

Equation (1) provides a way to comprehensively assess the performance of the gyroscope as a function of mechanical and magnetic properties.

The proposed magnetic readout approach can be envisioned to either replace the electrostatic readout scheme or to improve and assist the traditional electrostatic readout of capacitive-MEMS through a closed-loop scheme.

### B. Magneto-mechanical model

To design and characterize the device behavior, we consider for the magnetic proof mass $m$ the coupled equations of a two-degrees-of-freedom (2-DOF) vibratory gyroscope[30] along drive ($d$) and sense ($s$) directions:

$$\begin{cases} m\ddot{x} + c_{\mathrm{d}}\dot{x} + k_{\mathrm{d}}x = F_{\mathrm{drive}}(t) \\ m\ddot{y} + c_{\mathrm{s}}\dot{y} + k_{\mathrm{s}}y = -F_{c,y}(t) \end{cases} \tag{2}$$

where $c_{d(s)}$ is the viscous damping of the drive (sense) mode, $k_{d(s)}$ the elastic constant, related to the resonance frequency via $\omega_{d(s)} = \sqrt{k_{d(s)}/m}$, and linked to the quality factors $Q_{d(s)} = m\omega_{d(s)}/c_{d(s)}$, $F_{c,y}(t) = 2m\Omega_z(t)\dot{x}(t) = 2mX\omega_d\Omega_z(t)\cos(\omega_d t)$ is the Coriolis force acting on $m$ due to the angular rate $\Omega_{\mathrm{z}}$. The peak sense amplitude $Y$ induced by the Coriolis response is given by:

$$Y = \frac{2X\omega_d\Omega_{\mathrm{z}}(t)}{\sqrt{(\omega_s^2 - \omega_d^2)^2 + \left(\frac{\omega_s\omega_d}{Q_s}\right)}} \tag{3}$$

As in conventional MEMS gyroscopes[2,4], working in a matched mode ($\omega_s \approx \omega_d$) maximizes the sense response, and, under a steady-state sinusoidal drive ($\Omega_z(t) = \Omega_z$), $Y$ becomes:

$$Y \approx \frac{2Q_s}{\omega_d}X\Omega_z \tag{4}$$

where $Y \propto \Omega_z$ with a proportionality that scales with $Q_s$ and the drive amplitude $X$.

By linking Eq. 4 with the output signal to the displacement, we can express $\Delta V_{\mathrm{dc}} = \left(\frac{4SGXQ_s}{\omega_d}\right)\Omega_z = K_\Omega\Omega_z$, where $K_\Omega \equiv \partial\Delta \mathrm{V}_{\mathrm{dc}}/\partial\Omega_z = 4SGXQ_s/\omega_d$ represents the rate-to-voltage sensitivity and links mechanical design properties ($X$, $\omega_d$, $Q_s$) to magnetic readout properties ($S$, $G$). Therefore, $K_\Omega$ can be improved by increasing STD sensitivity $S$, field gradient $G$ (reduced spacing / larger magnetic moment), drive amplitude $X$, and $Q_s$ (packaging). We wish to underline that these parameters are not independent. $G$ can be increased by reducing the MTJs-proof mass distance or by increasing the proof mass magnetization at the cost of tighter fabrication tolerances. $X$ is limited by mechanical properties and the applicable actuation force. Increasing $Q_s$ enhances the matched-mode response while decreasing $BW$ and increasing the sensitivity to detuning and temperature drift effects, fabrication variations and mechanical imperfections. Therefore, proper optimization requires a dedicated multiphysics analysis beyond the scope of this proof-of-concept work.

It can be useful to transform the coupled equations in Eq. 2 in terms of non-dimensional parameters. This allows us to highlight the role of the mechanical damping and detuning in controlling the mechanical gain, defined as the ratio between the peak displacements $Y$ and $X$. By defining the normalized displacement along the sense-axis $y' = y/X$, introducing the mode separation parameter $\varrho = \omega_s/\omega_d$ describing the separation of the frequencies between drive and sense axes, the damping ratio $\zeta_s = 1/(2Q_s)$ and the detuning parameter $\Gamma = \Omega_z/\omega_d$, the sense response becomes:

$$\ddot{y}' + 2\zeta_s \varrho \dot{y}' + \varrho^2 y' = 2\Gamma \cos(\omega_d t) \quad (5)$$

Eq. 4 shows that, for perfect match-mode operation ($\varrho = 1$), the mechanical gain $|Y|/X$ is $\Gamma Q_s$ and depends only on the quality factor of the sense mode and the ratio between angular velocity and drive frequency. This gives a recipe to optimize $Y$ from a mechanical perspective.

The magnetization dynamics of the free layer of the two MTJs ($\boldsymbol{m}_1, \boldsymbol{m}_2$) and the magnetization of the magnetic proof mass ($\boldsymbol{m}_3$) are modelled (with macrospin equations derived from a micromagnetic approximation) [17,18]. This allows us to describe their dynamics via a set of three coupled Landau-Lifshitz-Gilbert (LLG) equations of motion [31,32]:

$$
\begin{cases}
\frac{d\boldsymbol{m}_1}{dt} = -\frac{\gamma_0}{1+\alpha_G^2}\left[\left(\boldsymbol{m}_1 \times \boldsymbol{b}_{\mathrm{eff},1}\right) + \alpha_G \boldsymbol{m} \right. \\
\qquad\qquad\qquad\qquad +\boldsymbol{\tau}_{\mathrm{STT},1} \\
\frac{d\boldsymbol{m}_2}{dt} = -\frac{\gamma_0}{1+\alpha_G^2}\left[\left(\boldsymbol{m}_2 \times \boldsymbol{b}_{\mathrm{eff},2}\right) + \alpha_G \boldsymbol{m}\right. \\
\qquad\qquad\qquad\qquad +\boldsymbol{\tau}_{\mathrm{STT},2} \\
\frac{d\boldsymbol{m}_3}{dt} = -\frac{\gamma_0}{1+\alpha_G^2}\left[\left(\boldsymbol{m}_3 \times \boldsymbol{b}_{\mathrm{eff},3}\right) + \alpha_G \boldsymbol{m}\right.
\end{cases} \quad (6)
$$

where $\boldsymbol{m}_i = \frac{\boldsymbol{M}_i}{M_{s,i}}$ $(i = 1,2,3)$ is the normalized magnetization vector of magnetic element $i$ with saturation magnetization $M_{s,i}$ and volume $V_i$, $\alpha_G$ is the Gilbert damping constant, $\gamma_0$ is the electron gyromagnetic ratio, and $t$ is the time. The effective field ($\boldsymbol{b}_{\mathrm{eff},i}$) includes a uniaxial anisotropy ($\boldsymbol{b}_{k,i}$) field, a demagnetization field ($\boldsymbol{b}_{\mathrm{dmag},i}$) and the dipolar field ($\boldsymbol{b}_{\mathrm{dip},j\to i}$) exerted by $j$ on $i$. The latter provides the magnetic coupling mechanism in the proposed device and is given by: $\boldsymbol{b}_{\mathrm{dip},j\to i} = \left(\mu_0/4\pi r_{ij}^3 \mu_0 M_{s,i}\right)\left[3\hat{\boldsymbol{r}}_{ij}\left(\boldsymbol{m}_j \cdot \hat{\boldsymbol{r}}_{ij}\right) - \boldsymbol{m}_j\right] = \overline{\overline{D}} \cdot \boldsymbol{m}_j$, where $\overline{\overline{D}}$ is the dipolar tensor. This tensor depends only on the distance between $i$ and $j$ seen as point-like objects, and $\hat{\boldsymbol{r}}_{ij} = \frac{\boldsymbol{r}_{ij}}{r_{ij}}$ is the distance unit vector. $\boldsymbol{\tau}_{\mathrm{STT},i} = \frac{g\mu_B\eta_T}{(e\gamma_0 M_{s,i} V_i)} J_{\mathrm{ac}}[(\boldsymbol{m}_i \times \boldsymbol{p}) - \alpha_G \boldsymbol{m}_i \times (\boldsymbol{m}_i \times \boldsymbol{p})]$ is the spin transfer torque (STT) acting on the STD $i$ $(i = 1,2)$ induced by the injected radiofrequency current $J_{\mathrm{ac}} = J_0 \sin(2\pi f_{ac})$ of frequency $f_{ac} = \frac{\omega_{\mathrm{ac}}}{2\pi}$ and amplitude $J_0$. $\eta_{\mathrm{T}} = 2\eta/(1+\eta^2 cos\vartheta)$ is the spin-polarization function and depends on the angle formed between $\boldsymbol{m}_i$ and the pinned layer (PL) magnetization ($\boldsymbol{p}$), and the spin-polarization efficiency $\eta$ [33–35].

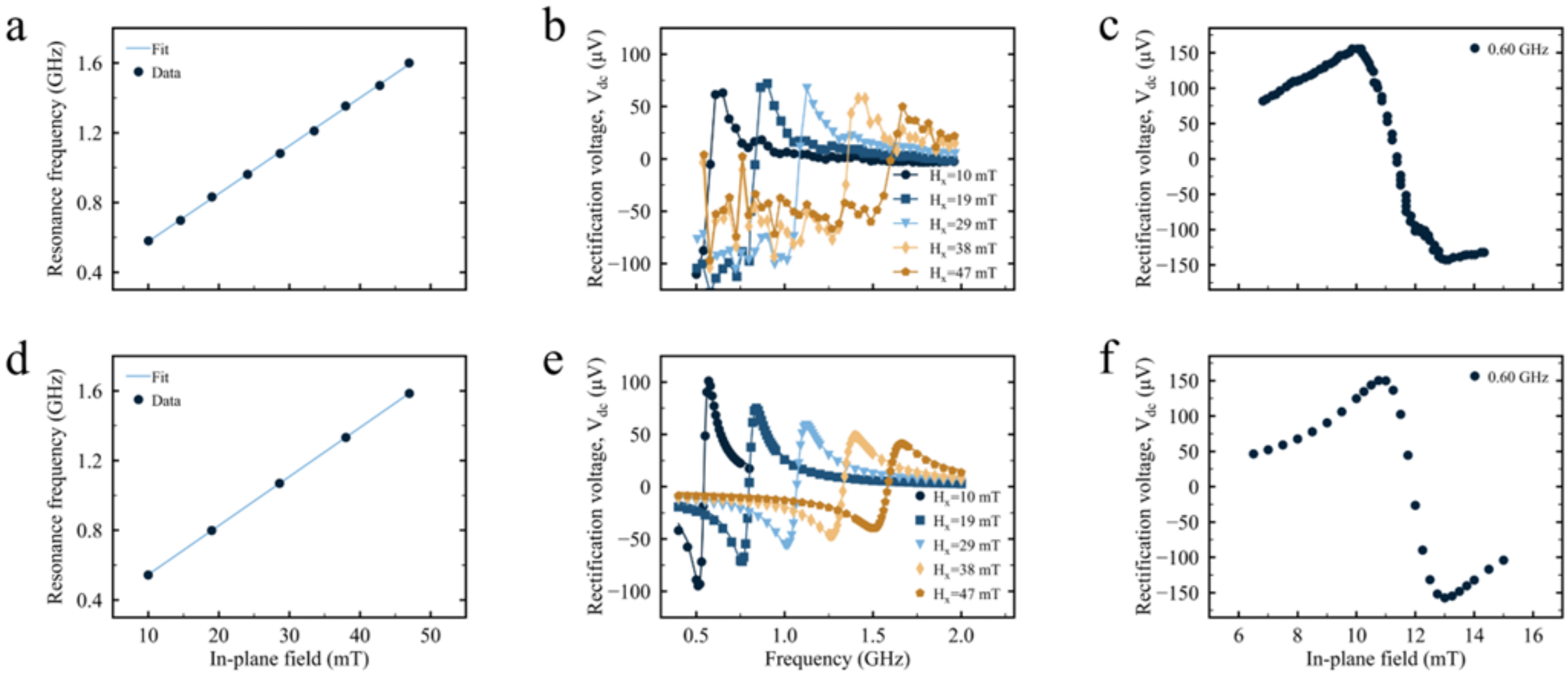

Fig. 2 (a) Comparison between experimental and simulation spin-torque diode characterization. (a,d) Resonance frequency dependence on in-plane applied field, showing the expected behavior for an in-plane magnetized system. (b,e) Frequency dependence of the rectification voltage $V_{\mathrm{dc}}$ generated by the MTJ via STD effect under different in-plane applied fields. (c-f) $V_{\mathrm{dc}}$ dependence on the in-plane applied field for $f_{\mathrm{ac}} = 0.60$ GHz.

## III. EXPERIMENTAL MAGNETIC CHARACTERIZATION

The identification of model parameters for the magnetic design is based on experimentally characterized MTJs with the following structure: Ta(6)/Ru(10)/Ta(10)/ CoFeB(0.85)/MgO(2)/CoFeB(1.5)/Ta(10)/Ru(20), patterned into elliptical $300\ \mathrm{nm} \times 750\ \mathrm{nm}$ cylinders. Thickness values are given in nm. We exploit MTJs in a hybrid configuration where the pinned layer (PL) is out-of-plane magnetized while the free layer (FL) is in-plane magnetized. This configuration maximizes the precessional cone and, hence, the rectified output voltage. Figure 2a shows the Kittel-like dependence of the resonance frequency ($f_{\mathrm{res}}$) as a function of the in-plane field applied along the long-axis (x-axis) of the MTJ. $f_{\mathrm{res}}$ ranges between 0.6 and 1.6 GHz for fields between 10 and 50 mT. Figure 2b presents STD measurements performed by applying microwave signals of $-10$ dBm power from 0.5 to 2 GHz modulated by a 101 Hz sinusoidal signal by a lock-in amplifier, and shows the frequency dependence of the rectification voltage $V_{\mathrm{dc}}$. These results highlight a μV amplitude of the output voltage, therefore allowing for readout applications. Finally, Figure 2c shows $V_{\mathrm{dc}}$ as a function of the external field, in the range of 6 to 15 mT, obtained by applying microwave signals of $-4$ dBm power at a frequency of 0.6 GHz, from which we can estimate a sensitivity $S{\sim}20$ μV/mT for these MTJs.

**Table I**. Model parameters.

| **Parameter** | **Value** | **Unit** |
|---|---|---|
| *Experimentally derived MTJ parameters* | | |
| Saturation magnetization, $M_s$ | $1.05 \times 10^6$ | A/m |
| Magnetic anisotropy energy density, $K_U$ | $0.68 \times 10^6$ | J/m$^3$ |
| Tunnel magnetoresistance, *TMR* | 15 | % |
| High resistance state, $R_{ap}$ | $34 \times 10^3$ | Ω |
| Low resistance state, $R_p$ | $28 \times 10^3$ | Ω |
| MTJ short axis length | 300 | nm |
| MTJ long axis length | 750 | nm |
| FL thickness $d$ | 1.5 | nm |
| *Estimated mechanical parameters* | | |
| Natural frequency sense mode, $\omega_s/2\pi$ | $250 \times 10^3$ | Hz |
| Viscous damping sense mode, $c_s$ | $1.0 \times 10^{-2}$ | kg s$^{-1}$ |
| Quality factor sense mode, $Q_s$ | $1.57 \times 10^2$ | |
| Bandwidth sense mode, *BW* | $1.592 \times 10^3$ | Hz |
| Proof mass, $m$ | $1 \times 10^{-6}$ | kg |

We use the magnetic parameters extracted from the experimental data, summarized in the top part of Table I, to parametrize the macrospin model (see section II.B), which we have first validated micromagnetically to ensure a uniform magnetic texture and a coherent magnetization dynamics. In addition, we consider $\eta = 0.66$ and $\alpha_{\mathrm{G}} = 0.02$[17,18], typical values for such MTJs. Figure 2d shows the simulated $f_{\mathrm{res}}$ as a function of the in-plane field in the same range of the experiments. Figure 2e reproduces the experimental $V_{\mathrm{dc}}$ frequency dependence for an injected current density $J_0 = 2.1 \times 10^3$ Am$^{-2}$, and Figure 2f shows that we successfully obtain the same field dependence of the experimental characterization, for $J_0 = 3.6 \times 10^3$ Am$^{-2}$. These results demonstrate excellent agreement with the experiments, allowing an experiment-informed design. In the rest of the paper, unless differently stated, the magnetic parameters and magnetic characterization discussed in this paragraph will be used.

## IV. RESULTS

In practical vibratory gyroscopes the drive axis is operated in a closed-loop amplitude-control scheme that maintains a constant drive displacement $X$ at $\omega_d$ despite variations in damping and quality factor. Accordingly, here we treat $X$ as closed-loop controlled, and, hence, it does not impact the reported $K_\Omega$, as long as the drive loop can supply the required actuation force to sustain $X$.

### A. Mechanical setup

We evaluate the coupled magneto-mechanical gyroscope model using the experimentally calibrated STD sensitivity $S$ from Figure 2(c). The simulated device consists of a magnetic proof mass $m$ elastically suspended between anchors along the drive-axis and magnetically coupled to two symmetrically placed STDs along the sense-axis at a nominal lateral spacing of 1 µm from the proof mass center. Unless otherwise stated, the gyroscope is operated in matched mode ($\omega_d \sim \omega_s = 2\pi \times 250.4$ kHz) to maximize the Coriolis sense response, with an open-loop bandwidth sense mode $BW \sim 1.6$ kHz. The mechanical parameters utilized to model the gyroscope are summarized in the bottom part of Table I. We note that, differently from the magnetic properties that are extracted from experimental characterization of the MTJs, the mechanical parameters are estimated[36–38] under the assumption of a total proof mass of $1.0 \times 10^{-6}$ kg and sense resonance frequency ~250 kHz. We underline that we consider the drive amplitude $X$ regulated by closed-loop control, which enforces the drive amplitude. We also wish to highlight that there exist approaches based on machine learning[39] and neural networks[40] that hat can support the discovery of material and structural parameters.

## B. Steady-state scenario

To demonstrate the operation of the proposed magneto-mechanical gyroscope, we calibrate the device by applying different steady-state $\Omega_z$ and extract the STDs differential $V_{\mathrm{dc}}$ ($\Delta V_{\mathrm{dc}} = V_{\mathrm{dc},1} - V_{\mathrm{dc},2}$) readout signal. Figure 3a shows $V_{\mathrm{dc}}$ versus the peak sense displacement $Y$ for several $\Omega_z$ values, confirming that the differential STD output increases monotonically with sense motion, and that the response remains approximately linear over the simulated range. Figure 3b plots $V_{\mathrm{dc}}$ and $Y$ as a function of $\Omega_z$ where it can be observed that both are well fit by linear relationships, yielding $K_\Omega =$ 0.0563 μV/(rad/s) $=$ 0.001 μV/(°/s) for this design. We wish to highlight that this value corresponds to an unoptimized design. Improving ARW requires increasing the intrinsic transduction gain $K_\Omega$ (via $S$, $G$, $X$ and $Q_s$) and/or reducing input-referred noise. Among others, possible pathways are optimization of the MTJs locations to enhance the sensitivity, optimization of the MTJ stack to enhance the rectification voltage output power and sensitivity of the STD effect.

Using the extracted $K_\Omega$, the MTJ resistance values, and the open-loop bandwidth, we next estimate standard inertial performance metrics to benchmark the present design point

## C. Metrics estimation

Using the measured MTJ resistances, we can estimate the MTJ Johnson noise density[23] $e_{n,J}(R) = \sqrt{4k_B TR}$, where $k_B$ is Boltzmann constant, $T$ = 300 K is the temperature and $R = \left(R_{ap} + R_p\right)/2$ is the average MTJ resistance. Considering a typical amplifier noise floor $e_{n,\mathrm{amp}} = 5\ \mathrm{n}V/\sqrt{Hz}$, the differential Johnson-noise-limited baseband noise density can be obtained as $e_{n,\Delta} = \sqrt{e_{n,J}^2\left(R_{MTJ}\right) + e_{n,\mathrm{amp}}^2} \approx 30\mathrm{n}V/\sqrt{Hz}$. By combining it with the estimated (pre-amplification) $K_\Omega =$ $0.001\ \mathrm{\mu}V/(°/\mathrm{s})$, we can estimate the angle random walk (ARW)[23] to be ARW $\approx \left(e_{n,\Delta}/K_\Omega\right) \cdot \sqrt{3600} = S_\Omega^{1/2} \cdot 60 = 1.8 \times 10^3$ °/√h, which is ~4/5 orders of magnitude larger than required for tactical lower bound (0.05 °/√h) and intermediate grade (0.005 °/√h)[1,41,42]. This value places the current unoptimized design in the rate-grade regime. Moreover, it highlights that improving ARW requires increasing the intrinsic transduction gain $K_\Omega$ rather than voltage gain alone, which would amplify signal and Johnson noise together.

The reported ARW reflects the white Johnson-plus-amplifier floor. As the rate signal is recovered by synchronous demodulation at $f_0$ = 250 kHz, only noise within the detection band around $f_0$ is translated to baseband; the MTJ 1/f resistance noise, concentrated near DC, lies outside this band and is rejected by the post-demodulation low-pass stage rather than appearing at baseband, suppressing its dominant contribution. MTJs are known to exhibit significant low-frequency 1/f resistance noise. Residual flicker entering as amplitude/phase modulation of the carrier, together with bias instability, to which 1/f noise primarily contributes, is not quantified here and is left to experimental Allan-variance characterization of fabricated devices, which represents the focus of future work.

### D. Quantitative comparison with capacitive readout

We separate the mechanical Brownian limit, common to both readout schemes, from the readout-added electronic noise. The Brownian rate-noise floor $S_\Omega^{1/2} \approx 1/X\sqrt{k_B T/m\omega_s Q_s}$ gives $\mathrm{ARW_{th}} \approx 0.04\,°/\sqrt{\mathrm{h}}$, the shared lower bound on the present resonator.

For the MTJ readout, the differential Johnson-plus-amplifier density $e_{n,\Delta} \approx 30\ \mathrm{nV}/\sqrt{\mathrm{Hz}}$, referred to rate through $K_\Omega = 0.001\ \mu\mathrm{V}/\,°/\mathrm{s}$, yields $S_{\Omega,\mathrm{MTJ}}^{1/2} = \frac{e_{n,D}}{K_\Omega} = 30\ (°/\mathrm{s})/\sqrt{\mathrm{Hz}}$, i.e. $\mathrm{ARW_{MTJ}} \approx 1.8 \times 10^3\,°/\sqrt{\mathrm{h}}$, which is about four and a half orders above the Brownian floor. To compare with capacitive sensing under the same mechanical rate-to-displacement conversion ($\frac{\partial x}{\partial \Omega} = \frac{K_\Omega}{2SG} \approx 1.25\ \mathrm{pm}/(°/\mathrm{s})$ for the present design), we refer to capacitive readout-to-rate in the literature. Ding *et al*. [43] report, for a switch-bridge readout applied to a capacitive MEMS gyroscope, a gain of 1.96 V/µm and a capacitance-noise density of $0.077\ \mathrm{aF}/\sqrt{\mathrm{Hz}}$, corresponding to an input-referred displacement noise of $\approx 0.28\ \mathrm{pm}/\sqrt{\mathrm{Hz}}$; through the same conversion this gives $\mathrm{ARW_{cap}} \approx 13\,°/\sqrt{\mathrm{h}}$. Thus, under identical rate-to-displacement conditions, the present MTJ readout is ~130x lower than this representative capacitive readout. Both readouts remain electronics-limited above the $0.04\,°/\sqrt{\mathrm{h}}$

Brownian floor, because the present low-Q resonator yields a weak rate-to-displacement conversion. Reaching the floor would require displacement resolution below ~1 $\text{fm}/\sqrt{\text{Hz}}$, beyond even state-of-the-art capacitive front-ends on this resonator. The MTJ-capacitive gap itself is set by the low displacement-to-voltage transduction gain ($2SG \approx 0.8\ \text{mV}/\mu\text{m} \approx 10^3$ below capacitive front-ends), i.e. by $K_\Omega$, not by the resonator.

For context, optimized capacitive gyroscopes report ARW down to $0.008\,°/\sqrt{\text{h}}$ [44] for higher-Q resonators and for which Brownian floors are correspondingly lower. This shows how the readout electronics is not the main bottleneck of these designs. Hence, the performance gap between the proposed solution and the established capacitive ones can be reduced through proper tuning of S, G, X, and $Q_s$.

### E. Real-case scenario

We next evaluate a time-varying angular rate $\Omega_z(t)$, representative of a realistic scenario where a drone suffers from a wind gust. Figure 3c shows the time-trace of the considered $\Omega_z(t)$, modulated at 40 kHz with an average amplitude of 40 rad/s. Figure 3d presents the one-to-one mapping of $\Delta V_{dc}(t)$ with $Y(t)$. The results show how the simulated $\Delta V_{dc}(t)$ contains a carrier at the mechanical drive frequency, whose amplitude is slowly modulated by $\Omega_z(t)$ through the sense displacement.

To recover $\Omega_z(t)$ from the measured voltage, we perform digital dual-phase (I/Q) homodyne demodulation at the known drive frequency $\omega_d$ extracting the in-phase ($I(t)$) and quadrature ($Q(t)$) components of the signal, and applying a lock-in based demodulation (see Appendix B for more details). Figure 3e compares $I(t)$ and $Q(t)$ highlighting how, through this approach, $Q(t)$ is about three orders of magnitude smaller than $I(t)$ and, finally, Figure 3f presents the normalized error $100 \times \left|\Omega_z^{\text{Input}} - \Omega_z^{\text{Measured}}\right| / \Omega_z^{\text{Input}} = \Delta\Omega_z / \Omega_z^{\text{Input}}$ in the extraction of $\Omega_z(t)$ obtained from the demodulation process. The error is ∼0.05 % across with maximum of ∼0.35% at the edges, due to signal alignment, and where the strongest variations in $\Omega_z(t)$. Nevertheless, the agreement is excellent and demonstrates that the proposed magneto-mechanical gyroscope is able to extract successfully the angular velocity in realistic scenarios ($\Omega_z(t)$).

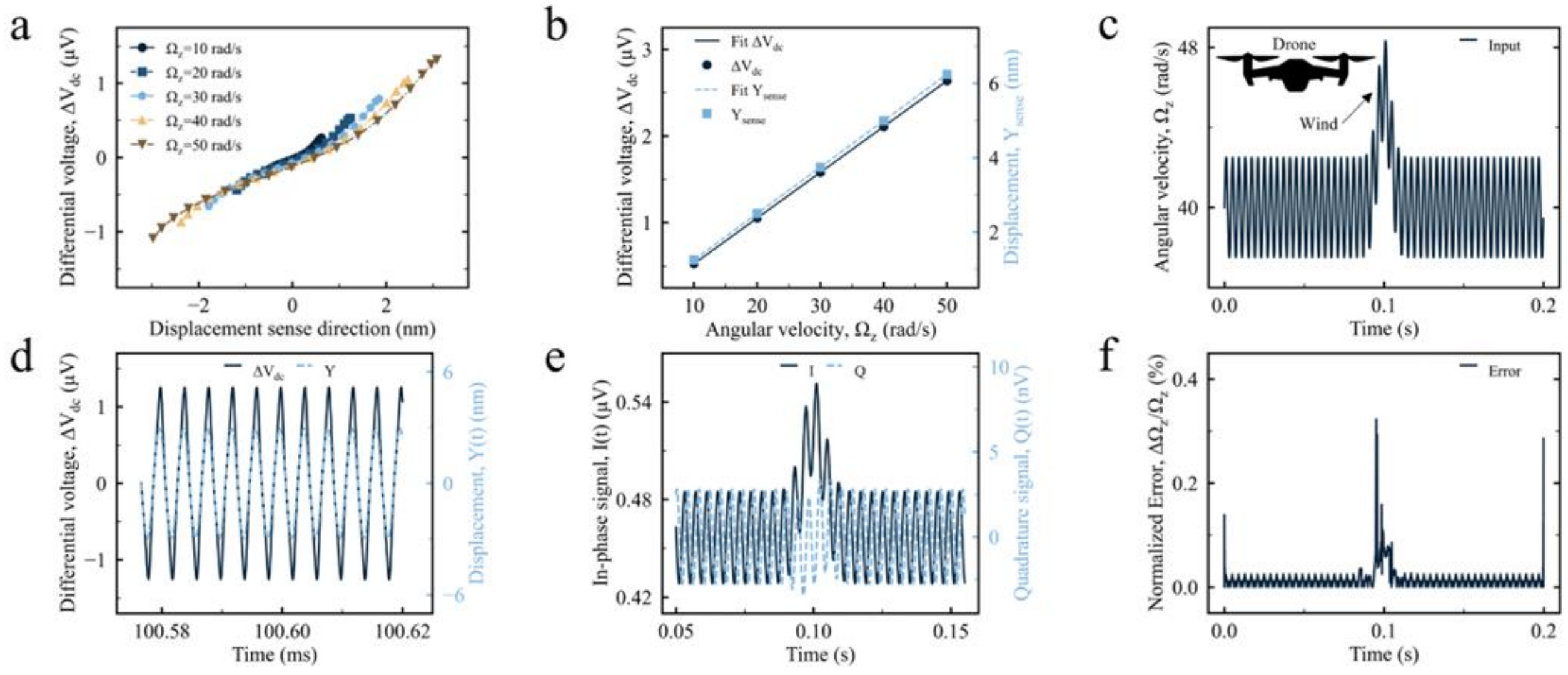

Fig. 3 (a) Dependence of the differential output voltage ($\Delta V_{\mathrm{dc}}$) on the magnetic proof mass displacement along the sense direction for steady-state $\Omega_z = 10, 20, 30, 40, 50$ rad/s. (b) Linear dependence of $\Delta V_{\mathrm{dc}}$ (black dots) and peak amplitude of magnetic proof mass displacement along sense direction ($Y_{\mathrm{sense}}$, light-blue squares) on $\Omega_z$. (c) Input time-dependent profile of $\Omega_z(t)$ and (d) one-to-one mapping of $\Delta V_{\mathrm{dc}}$ and displacement along the sense direction ($Y$) as a function of time. (e) Zoom of the in-phase ($I$) and quadrature ($Q$) components of the signal extracted using a filtering algorithm. (f) Normalized error $100 \times \left|\Omega_z^{\mathrm{Input}} - \Omega_z^{\mathrm{Measured}}\right| / \Omega_z^{\mathrm{Input}} = \Delta\Omega_z / \Omega_z^{\mathrm{Input}}$ in the extraction of $\Omega_z(t)$ obtained from the demodulation process.

### F. A full-spintronic design

As a design variant, the magnetic proof mass can be replaced by a spin-torque nano-oscillator (STNO) that can be realized with the same MTJ, now driven also by a dc current, analogous to prior spintronic magneto-mechanical accelerometer concepts[17,18]. In this configuration the STNO acts as an on-chip RF source whose oscillation can magnetically couple dynamically to one or more nearby STDs. This synchronization among STDs and STNO, together with the injection locking phenomenon [11,28,34,45,46], offers a path to increased sensitivity, at the cost of tighter control of MTJ parameters and spacing.

To verify the operation of this alternative design, we study the response under a steady-state $\Omega_z$, similarly to Section IV.C. We highlight that $X$ is limited to 150 nm for the device to successfully operate, yielding a potential $3\times$ reduction in dimensions along the drive axis. The magnetic and mechanical parameters adopted for this solution, and more details, are presented in Appendix C. Figure 4 shows the linear dependence of $\Delta V_{\mathrm{dc}}$ and peak $Y$ with $\Omega_z$, from which we extract $K_\Omega = 3.75$ μV/(rad/s) $= 0.065$ μV/(°/s), a value $\sim 70\times K_\Omega$ of the original gyroscope design and that yields ARW $\sim 25\ °/\sqrt{h}$. While this ARW maintains the proposed solution in the rate-grade regime, these results demonstrate the potential for improvement of magneto-mechanical spintronic-based gyroscopes, in both performance and dimensional scaling.

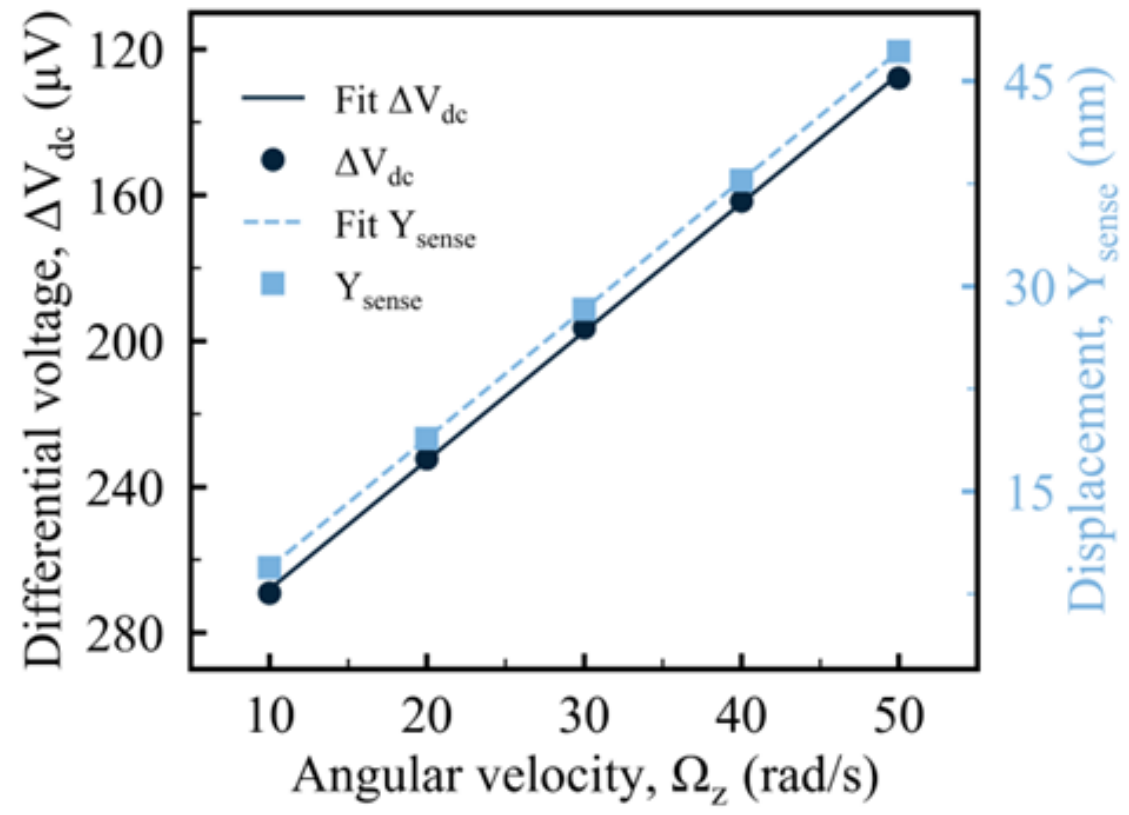

Fig. 4 Linear dependence of $\Delta V_{\mathrm{dc}}$ (black dots) and peak amplitude of the STNO (proof mass) displacement along the sense direction ($Y_{\mathrm{sense}}$, light-blue squares) on $\Omega_z$.

## V. CONCLUSIONS

We introduced a model for a spintronics-based vibratory gyroscope in which the sense displacement of a proof mass is converted into a DC voltage by MTJs operated as STDs. The key engineering feature of this architecture is a direct displacement-to-voltage readout scheme that can replace or improve the capacitance-to-voltage method used in MEMS gyroscopes, while preserving the traditional drive/sense mechanical paradigm and synchronous demodulation methods. Using experimentally characterized STD responses to calibrate a coupled magneto-mechanical model under matched-mode operation, we demonstrated linear rate response in simulation for both steady-state and time-varying inputs, integrating a fully-developed filtering algorithm to extract the actual measured signal. This solution is scalable to a configuration using active STD and STNO that offers further dimensional scalability and potential performance improvements. Therefore, the results of this work highlight a multi-physics platform of magneto-mechanical system integrated with spintronic technology as a promising platform for next-generation gyroscopic sensors.

## VI. APPENDIX

### A. Stray field evaluation

The proposed readout scheme relies critically on the change of the stray field distribution upon displacement of the proof mass. To demonstrate this, we have extracted the stray field spatial distribution from three different configurations obtained during the modelling of the gyroscope upon application of a static $\Omega_z = 40$ rad/s presented in Fig.3a: (a) $X = -350$ nm, $Y = +2$ nm (b) $X = 0$ nm, $Y = 0$ nm and (c) $X = +350$ nm, $Y = -2$ nm. These are plotted in Fig. A1. These results show how the stray field distribution differs depending on the configurations, demonstrating the core principle of the proposed readout scheme.

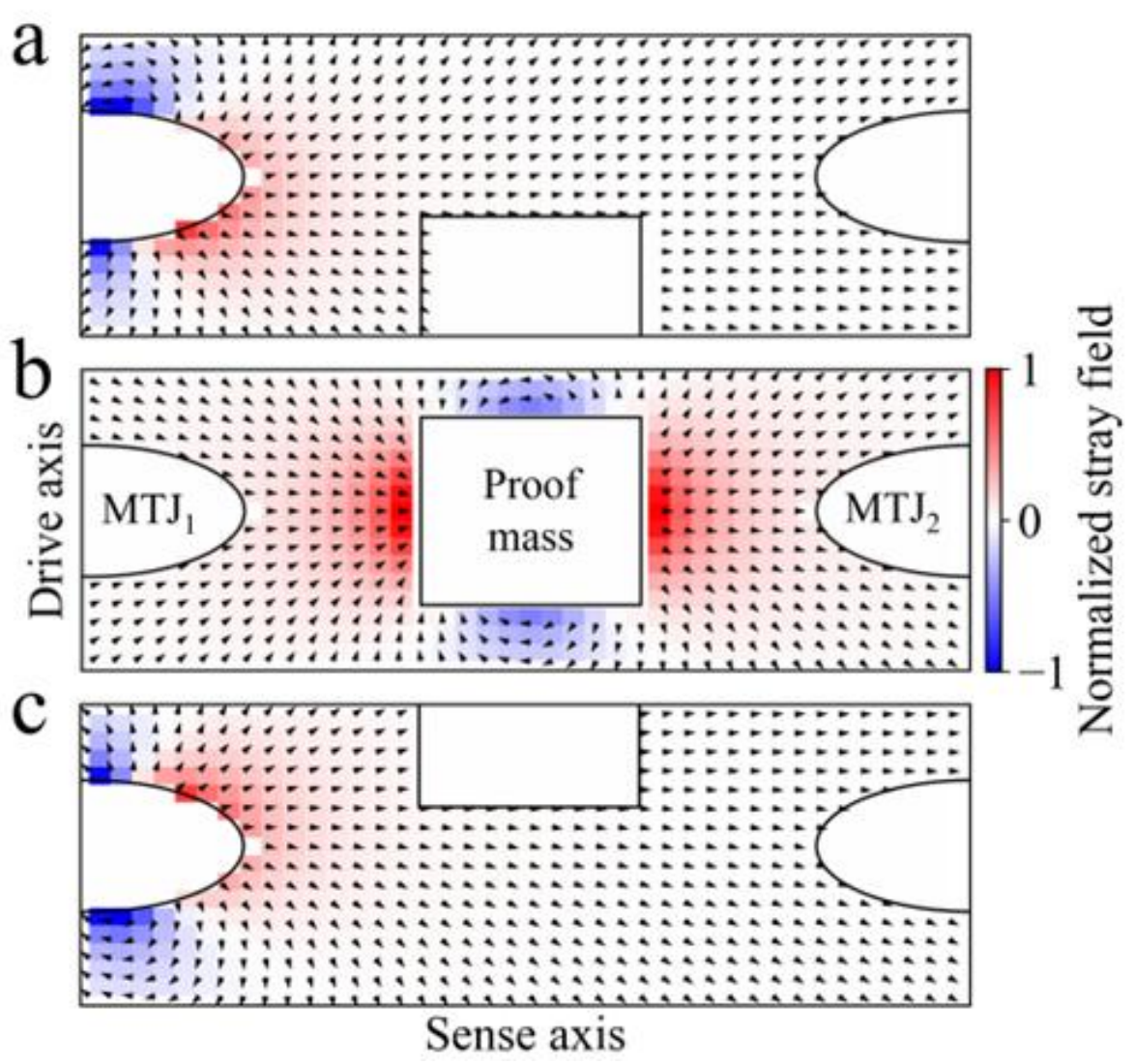

Fig. A1 Stray field spatial distribution. Plot of the normalized stray field maps corresponding to different displacements of the proof mass from the central position: (a) $X = -350$ nm, $Y = +2$ nm (b) $X = 0$ nm, $Y = 0$ nm and (c) $X = +350$ nm, $Y = -2$ nm. The colour palette represents the component of the stray field along the sense axis, while arrows give the in-plane direction of the field.

### B. Demodulation algorithm

Angular-rate information was extracted from the simulated differential voltage $\Delta V(t)$ using digital complex lock-in (homodyne) demodulation followed by an effective linear calibration. Figure A2 presents the workflow of the demodulation.

After subtracting the time average $\langle \Delta V_{dc} \rangle$, the signal was mixed with a complex reference at the drive frequency $f_0 = \frac{\omega_d}{2\pi}$: $z(t) = (\Delta V_{dc}(t) - \langle \Delta V_{dc} \rangle) e^{-j\omega_d t}$. $z(t)$ was then low-pass filtered to isolate the baseband using a 4th-order Butterworth filter with cutoff $f_c$:

$$\begin{cases} I(t) = LPF[(\Delta V_{dc}(t) - \langle \Delta V_{dc} \rangle) cos(\omega_d t)] \\ Q(t) = LPF[(\Delta V_{dc}(t) - \langle \Delta V_{dc} \rangle) sin(\omega_d t)] \end{cases} \quad \text{(A1)}$$

where $I(t)$ and $Q(t)$ are the in-phase and quadrature components. To suppress boundary transients without introducing group delay, the signal was mirror-padded and filtered using forward–backward (zero-phase) filtering, yielding the complex baseband signal $z_{\text{bb}}(t)$. A constant phase rotation $e^{-j\varphi_{\text{rot}}}$ was subsequently applied, with $\varphi_{\text{rot}}$ chosen to minimize quadrature power so that the demodulated observable is the in-phase component $I(t) = \Omega_{\text{LIA}}(t) = \Re\{z_{\text{bb}}(t) e^{-j\varphi_{\text{rot}}}\}$. $f_c$ was selected as $f_c =$ 3 kHz given that the informative baseband content is confined within a few hundred hertz, giving a margin of about an order-of-magnitude to avoid attenuating the baseband envelope while strongly rejecting the image around $2f_0$ and other high-frequency mixing products.

Since the electrical readout is governed by a nonlinear map $\Delta V_{dc}(x, y)$, but the baseband bandwidth ($\sim 10^2$Hz) is orders of magnitude smaller than the carrier ($f_0 = 2.5 \times 10^5$Hz), higher-order components are largely suppressed by the low-pass stage. This leaves an effective near-linear dependence of $\Omega_{\mathrm{LIA}}(t)$ on angular rate. Accordingly, $\Omega_{\mathrm{LIA}}(t)$ was aligned to the reference rate $\Omega_{\mathrm{true}}(t)$ obtained by least-squares fitting:

$$\Omega_{\mathrm{true}}(t) \approx a\Omega_{\mathrm{LIA}}(t) + b \qquad \text{(A2)}$$

giving $a = 8.8177 \times 10^{7}\,\mathrm{rad/s/V}$, $b = -0.264\ \mathrm{rad/s}$, with a residual RMSE $= 1.10 \times 10^{-2}$ rad/s.

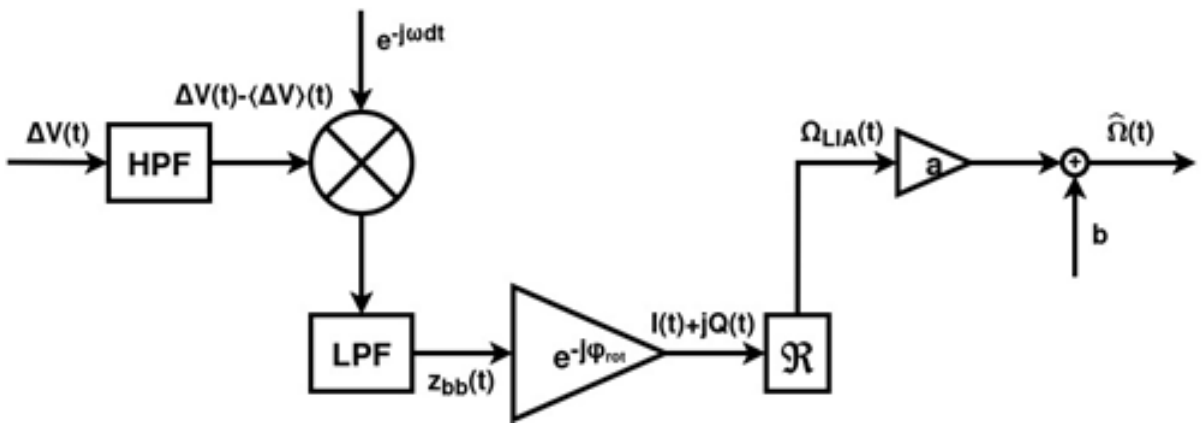

Fig. A2 Demodulation algorithm workflow.

### C. Spintronic solution with active STDs and STNO

In the full-spintronic design, we consider MTJs optimized for STD response and STNO behaviour, such as those widely described and employed in Refs.[28,34,45,46]. These MTJs have elliptical cross sections of dimensions 70 nm × 150 nm, a FL thickness of 1.63 nm, and the following magnetic parameters: $M_{\mathrm{s}} = 9.50 \times 10^{5}$ A/m, $K_{\mathrm{u}} = 5.45 \times 10^{5}$ J/m$^3$, $\alpha_G = 0.02$, $R_{AP} = 1.2$ kΩ and $R_P = 0.6$ kΩ. Given the reduced dimensions requirements imposed by the synchronization between STD and STNO, we consider the following mechanical parameters: $\omega_d \sim \omega_s/2\pi = 50$ kHz, $c_s = 3.14 \times 10^{-4}$ kg s$^{-1}$ and $Q_s = 1 \times 10^{3}$. Figure A3a shows a sketch of this design, where the magnetic proof mass is replaced by a STNO.

Figure A3b shows the dependence of $\Delta V_{\mathrm{dc}}$ on the STNO displacement as a function of different steady-state $\Omega_z$. $\Delta V_{\mathrm{dc}}$ exhibits a larger amplitude and a stronger variation, on the order of 70 µV, as well as a net change in the slope. In this design, as $\Omega_z$ increases, the STDs and the STNO are drawn further apart, reducing the magnetic coupling between them. This affects the synchronization, resulting in a reduction of $\Delta V_{\mathrm{dc}}$.

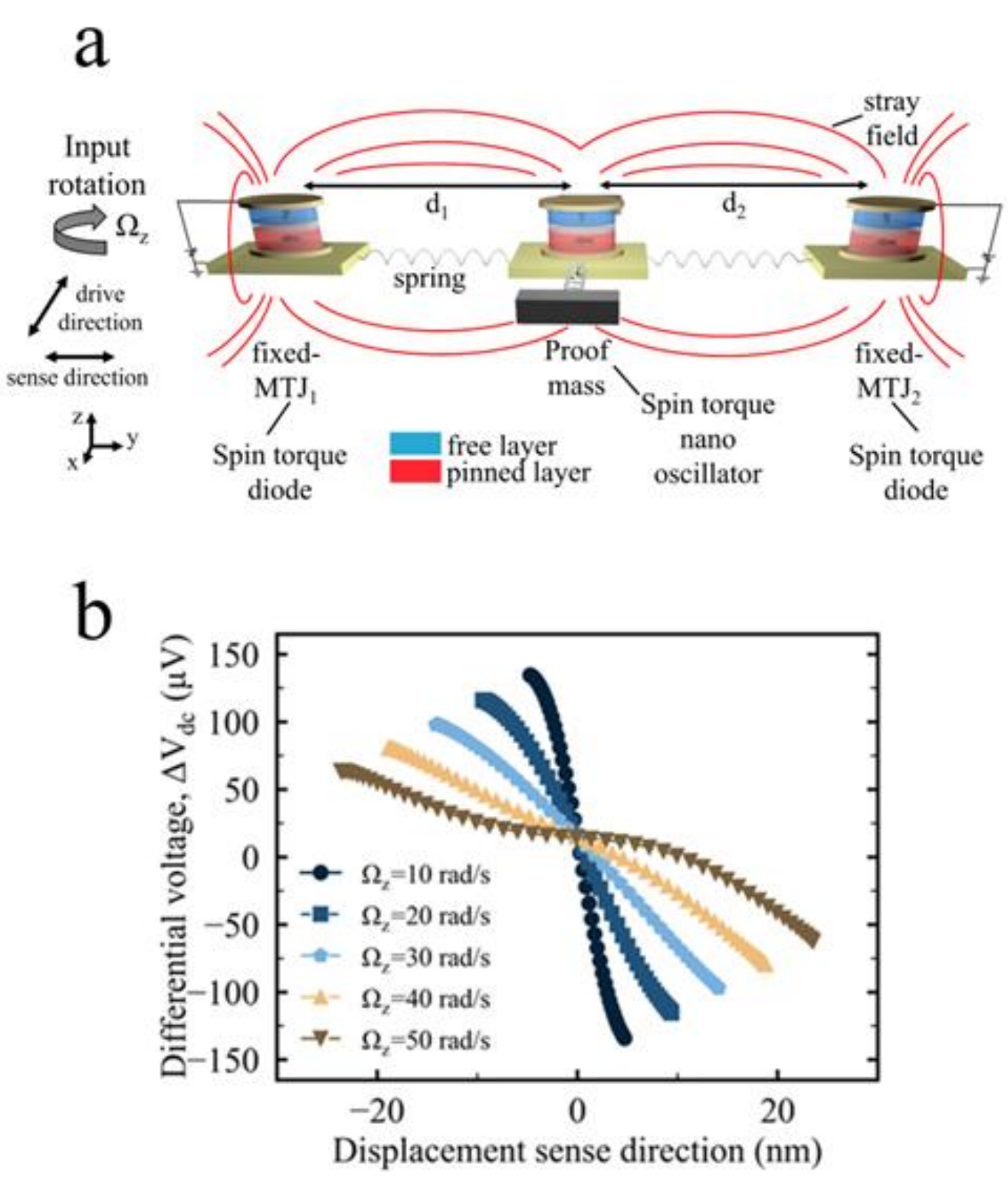

Fig. A3 (a) Sketch of the spintronic gyroscope based on active STDs and STNO. (b) Dependence of the differential output voltage ($\Delta V_{\mathrm{dc}}$) on the magnetic proof mass displacement along the sense direction for steady-state $\Omega_z = 10, 20, 30, 40, 50$ rad/s.

**ACKNOWLEDGEMENTS**

The work of A.M., M.F., R.T., M.C., F.G., G.F. was supported partially under the project PRIN_20225YF2S4 – Magneto-Mechanical Accelerometers, Gyroscopes and Computing based on nanoscale magnetic tunnel junctions (MMAGYC). A.M., R.T., M.C. and G.F also acknowledge support under the Contract n. 2025-40-I.0 (SPINAM) funded by the Italian Space Agency within the call "Studi di concetti innovativi di sistemi spaziali. P.K.A., V.L.D. and B.S. acknowledge support from the U.S. National Science Foundation under award number 2203242. A.M., M.F., R.T., M.C., F.G., G.F. are with the PETASPIN team and thank the PETASPIN association (www.petaspin.com).